\documentclass[epj]{svjour}

\usepackage{graphicx}
\usepackage{fancyhdr}
\def\makeheadbox{{
 }}
\def\mytitle{My title} 
\def\myauthors{My name}  
\def\mytype{My type of session}
\def\mysession{My session}

\def\mytitle{Search for GMSB NLSPs at LHC} 
\def\myauthors{Piotr Zalewski}    
\def\mytype{Contributed Talk}    
\def\mysession{Colliders - SUSY Phenomenology}

\newcommand{\pTmiss}{\mbox{$p_T^{\rm miss}$}}
\newcommand{\Meff}{\mbox{$M_{\rm eff}$}}
\newcommand{\ctau}{\mbox{$c\tau$}}
\newcommand{\ibeta}{\mbox{$\beta^{-1}$}}
\newcommand{\Mmm}{\mbox{$M_{\mu\mu}$}}
\newcommand{\Smajor}{\mbox{$S_{\rm major}$}}
\newcommand{\Sminor}{\mbox{$S_{\rm minor}$}}
\newcommand{\GeV}{\mbox{\rm GeV}}
\newcommand{\TeV}{\mbox{\rm TeV}}
\newcommand{\pTjth}{\mbox{$p_T^{4j}$}}
\newcommand{\pTgst}{\mbox{$p_T^{1\gamma}$}}

\newcommand{\pTm}{\mbox{$p_T^{\mu}$}}
\newcommand{\ra}{\rightarrow}
\def\t#1{\widetilde{#1}}
\newcommand{\stau}{\mbox{$\tilde{\tau}_1$}}
\newcommand{\slep}{\mbox{$\tilde{\ell}$}}
\newcommand{\nino}{\mbox{$\tilde{\chi}^0_1$}}

\newcommand{\EPS}{.}

\begin{document}
\title{Search for GMSB NLSPs at LHC}
\author{Piotr Zalewski\inst{1}
\thanks{\emph{Email:} piotr.zalewski@fuw.edu.pl}
\inst{2}
}                     
%
%
\institute{Soltan Institute for Nuclear Studies
\and on behalf of CMS and ATLAS collaborations
}
%
\date{}
\abstract{
NLSP -- LSP decays could have dramatic influence on SUSY
phenomenology at LHC. NLSP could have significant lifetime and could
be charged. In at least two scenarios
detectors must be used in a special way. They were not optimized
for detection of heavy (semi)stable charged particles 
and decaying in flight (neutral or charged) NLSPs.
During the last decade both ATLAS and CMS collaboration
have developed strategies which allow for effective search
within such scenarios.
} 

\maketitle
%

%
%

\section{Introduction}
\label{intro}

Within Gauge-Mediated ~Supersymmetry ~Breaking models gravitino
is the Lightest Supersymmetry Partner (LSP) whereas neutralino
or stau plays the role of the Next to Lightest Supersymmetry Partner (NLSP).
NLSP decays to its Standard Model partner and gravitino with a lifetime
depending on the scale of the SUSY breaking. Detection of the NLSP and
determination of its properties,
in particular lifetime, can be of crucial
importance for the physics program at LHC.

Minimal GMSB model is defined by six parameters:
$\Lambda$ -- effective SUSY mass scale, $N$ -- number of messenger
generations, $M$ -- messenger mass scale, $\tan{\beta}$,
$\rm sgn\,\mu$ and $C_g$ -- ratio of the intrinsic SUSY 
breaking scale to messenger SB
scale, governing goldstino coupling and hence -- NLSP lifetime.

\section{Long-lived charged NLSP at CMS}
\label{sec:st} 

For large $N$, typically right sleptons are the lightest supersymmetric
partners of SM particles. If, in addition,  $\tan{\beta}$
has not too low value, \stau\ is the NLSP which could have
arbitrary long lifetime. In this way one obtains benchmark scenario
in which in every supersymmetric event cascade decays ends on pair of
stable massive charged
and not strongly interacting particles. These particles propagate through
the detector like muons but with velocity $\beta$ smaller than 1.
Because of that their specific ionization is greater than for MIP
and they arrive at given detector layer with a time delay. Both facts
could be used to distinguish them from muons.

There are other theoretical scenarios with heavy stable charged particles.
ATLAS and CMS searches for them were reported by Shikma Bressler in the 
same session \cite{ref:bressler}. I will cover only some details 
concerning search for staus using TOF method in the CMS detector. The first
full detector simulation (based on GEANT3) analysis were performed 
already a~decade ago 
\cite{ref:zalewski98,ref:zalewski99}. 
However specific ionization were not correctly simulated then. 
Because of that the analysis was based on originally developed method of 
TOF measurement 
by drift tubes of the barrel muon system of the CMS.   

Here we present an update \cite{ref:zalewski06} of that analysis performed 
using OSCAR-ORCA version of CMS detector simulation 
(based on GEANT4), 
the same as used for CMS Physics TDR \cite{ref:PTDR}.

\subsection{The TOF method}
\label{sec:TOF}

The barrel muon system of the CMS detector consist of four concentric
muon stations inserted in the return yoke of the CMS solenoid. In each
station there are three super-layers (SL) of four layers of drift tubes (DT) 
each. Two of SL measure $R\phi$ coordinate and one $z$ coordinate (there is
no $z$ SL in the outermost station). The tubes in each SL are staggered by
half a~tube. An average track pass alternatively on left and right side
of the sensitive wire. 

In a given SL hits due to muon should align if
timing is correct whereas
hits due to delayed particle do not align,
they are shifted backward from the wire by $\delta_x$ and form a zig-zag 
pattern.
For each hit
$$
 \frac{{\delta_x}}{v_{\rm drift}}={\delta_t}=t_{\beta<1}-t_c=\frac{L}{c}(\frac{1}{\beta}-1)
$$
and hence measuring $\delta_x$ allows to estimate \ibeta 
$$
 \frac{1}{\beta}=1+\frac{{\delta_x}}{L}\frac{c}{v_{\rm drift}}.
$$
The exact formula used in the analysis to obtain \ibeta\ estimate
from the zig-zag pattern
is described elsewhere \cite{ref:zalewski06} but it could be 
simplified as follows:
$$
\frac{1}{\beta}=1+\frac{c}{v_{\rm drift}}\frac{1}{N}
               \sum_{i=1}^{N}\frac{{\delta_{x}^i}}{L_i}
$$
where $L_i$ is the flight distance, $v_{\rm drift}$ is the drift velocity,
${\delta_{x}^i}=|{x^{\rm hit}_i }-x^{\rm wire}_i|
               -|{x^{\rm reco}_i}-x^{\rm wire}_i|$ 
and $x^{\rm reco}_i$ is the local track element position at layer $i$.

\subsection{Results}
\label{sec:st-res}

Two points from the SPS7 line \cite{ref:SPS78}: $\Lambda=50\,\TeV$ and
$\Lambda=80\,\TeV$
were chosen for the full detector simulation. The mass of \stau\ particle
is equal to 152.31\,\GeV\ and 242.93\,\GeV\ respectively and the cross section
is 1\,pb and 0.1\,pb. In the analysis a high mass ($\Mmm>110GeV$) pair 
of energetic muons ($\pTm>60\,\GeV$) were searched for. Since most of the
signal events contains products of a cascade decays of squarks or gluinos 
a high effective mass of the event were required 
($\Meff>360\,\GeV$) as well. 
Taking this into account the following background sources
of muon pairs were considered: Drell-Yan above $Z^0$ resonance, $t\bar t$ and
double vector boson production. Events were triggered
on single muons with a default threshold of ($\pTm>80\,\GeV$).

\begin{figure}
\includegraphics[width=0.46\textwidth,height=0.30\textwidth,angle=0]%
{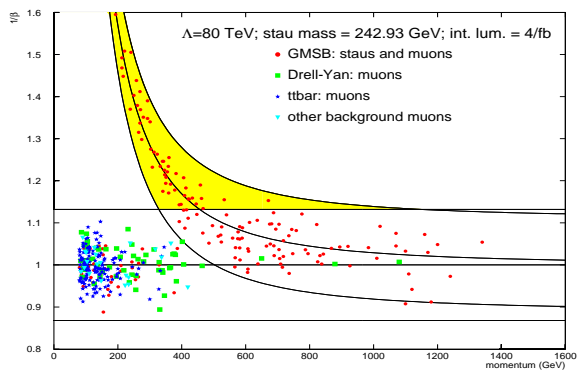}\\
\caption{Scatter plot \ibeta\ versus momentum}
\label{fig:st}
\end{figure}

Breakdown of number of events at different stages of the selection 
is given in the Table~\ref{tab:st}.  A scatter plot \ibeta\ versus momentum
for heavier mass point
and ${\cal L}=4/{\rm fb}$ 
is shown in the Figure~\ref{fig:st}
where clear separation of signal -- staus and background -- muons 
could be seen.
The last column in the Table~\ref{tab:st} correspond to highlighted area in 
that Figure.

\begin{table}
\caption{%
Number of events at different stages of the selection for 1/fb.
Columns correspond to: presel. -- trigger and $p_{\mu}>80\,\GeV$; 
quality -- additional requirements at the pattern recognition stage that 
significantly reduce
tails of \ibeta\ estimate 
(see ref.~\protect\cite{ref:zalewski06} for details); 
select. -- requirements described in the text; 
\ibeta\ -- highlighted area in the Figure~\protect\ref{fig:st}.}%
\label{tab:st}
\begin{tabular}{|l|rrr|r|}
\hline
dataset            & presel. & quality &   select. & \ibeta \\
\hline
$\Lambda=50\,\TeV$& 1714.1            &      956.4&  666.4&  155.054\\
$\Lambda=80\,\TeV$&  108.8            &       59.8&   45.0&   12.019\\
\hline
DY 2$\mu$ 
                       & 8105.6            &     4422.6&   13.6&    0.012\\
tt 2$\mu$                 & 2686.0            &     1624.4&   33.7&    0.029\\
WW 2$\mu$                 &  573.7            &      327.7&    6.0&    0.005\\
ZZ 2$\mu$                 &  202.0            &      110.1&    0.1&    0.000\\
ZW 2$\mu$                 &  231.6            &      121.3&    0.0&    0.000\\
\hline
$\Sigma$               &11798.9            &     6606.1&   53.4&    0.046\\
\hline
\end{tabular}
\end{table}

The upper limit for number of expected background events in the region 
above the highest
horizontal line in the figure ($\ibeta>1+3\sigma_{\beta^{-1}}=1.132$)
was evaluated to be 0.05 events for 1/fb which means that
the $5\sigma$ discovery could be
claimed with 8 signal events.
This correspond to integrated luminosity needed for a discovery
${\cal L}=52/{\rm pb}$ for $\Lambda=50\,TeV$ and
${\cal L}=667/{\rm pb}$ for $\Lambda=80\,TeV$.

It should be stressed, however, that real performance of the analysis
could not be evaluated without checking the method against muons
from real data.

Another question is how well the \stau\ mass could be determined.
To test this 
1000 pseudo-experiments corresponding to integrated luminosity
$
{\cal L}=0.5/{\rm fb}
$
for the lower mass point
(${\cal L}=4/{\rm fb}$ for the higher mass point)
were performed. 
The result are the following\\
$
M_{\stau}^{\rm est.}=
\{153.2\pm1.6(\rm stat.)\pm0.9(\rm syst.)\}\GeV
$\\
for generated mass
$
M_{\stau}^{\rm gen.}
=152.31\,\GeV
$
and\\ 
$
M_{\stau}^{\rm est.}=
\{243.2\pm3.2(\rm stat.)\pm1.4(\rm syst.)\}\GeV
$\\
for generated mass
$
M_{\stau}^{\rm gen.}
=242.93\,\GeV.
$
The test confirms that the method is well suited for the search
for stable staus. 
 
\section{Non-pointing photons in ATLAS} 
\label{sec:atlas}

Within GMSB 
if $N$ is small typically \nino\ is the NLSP. 
Its lifetime is
free parameter. By measuring lifetime it is possible to estimate
the scale of supersymmetry breaking which is otherwise not accessible
experimentally. 

In the ATLAS analysis~\cite{ref:atlas}
very interesting
technique was developed which allows
to determine the masses of the slepton and neutralino from events with
a~lepton and converted photon arising from the cascade decay
$\slep\ra\ell\nino\ra\ell\gamma\t{G}$.

\begin{figure}
\begin{center}
\includegraphics[width=0.35\textwidth,height=0.25\textwidth,angle=0]%
{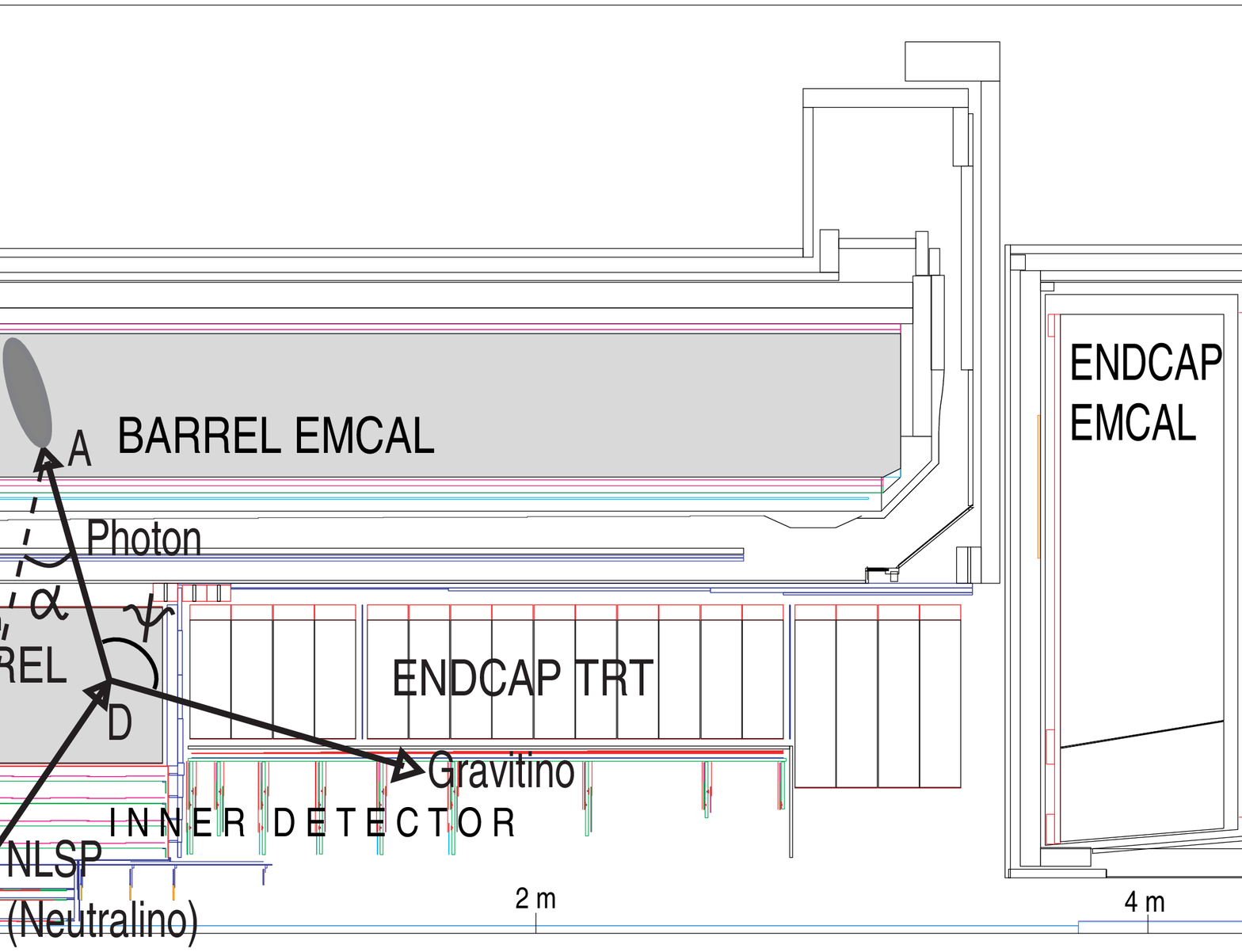}
\end{center}
\caption{The topology of \nino\ decay
inside ATLAS detector.}
\label{fig:atlas}
\end{figure}

The topology of \nino\ decay
inside ATLAS detector if the decay length is of the order of 1\,m is shown
in the Figure~\ref{fig:atlas}. 
In the case of photon conversion
the angle $\alpha$ could be precisely measured. ATLAS EMCAL allows
also for precise determination of photon arrival time $t_{\gamma}$. 
Since the distance $L=|\vec{OA}|$ is also known the angle $\psi$
could be determined~\cite{ref:atlas}
$$
\cos\psi =\frac{1-\xi^2}{1+\xi^2} ~~~{\rm where}~~~ 
\xi=\frac{ct_{\gamma}-L\cos\alpha}{L\sin\alpha}.
$$

If both angles $\alpha$ and $\psi$ are know for at least two reconstructed
$\slep\ra\ell\nino\ra\ell\gamma\t{G}$ cascade decays it is possible to
determine both $\t\ell$ and \nino\ masses.

Knowing these masses it is possible to analytically calculate
decay time and momentum
of neutralino using the ECAL data and lepton momentum only.
for each lepton non-pointing (converted or not) photon pair from slepton
decay.

This idea was tested with fast simulation of ATLAS detector for
GMSB point:
${\Lambda} = 90\,\TeV$, ${M} = 500\,\TeV$, ${N}=1$,
${\tan\beta}=5$, ${\mu}>0$.
Neutralino and slepton masses for this point are
$M(\nino)=117\,\GeV$, $M(\slep_R)=162\,\GeV$

\begin{figure}
\includegraphics[width=0.5\textwidth,height=0.24\textwidth,angle=0]%
{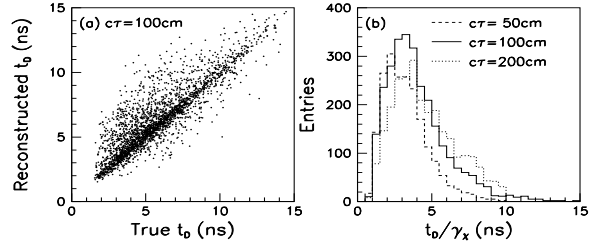}
\caption{Reliability of $t_D$ determination (a) 
and $t_D/\gamma_{\chi}$ distributions (b).}
\label{fig:at-ctauplot}
\end{figure}

\begin{figure}
\includegraphics[width=0.5\textwidth,height=0.24\textwidth,angle=0]%
{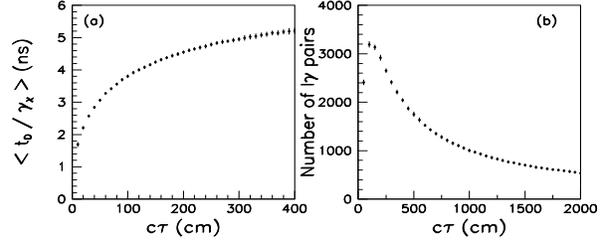}
\caption{Average $\langle t_D/\gamma_{\chi}\rangle$ (a) and
number of lepton non-pointing photon pairs (b).
}\label{fig:at-avedtg}
\end{figure}

Longitudinal segmentation of the EMCAL of ATLAS allows for precise
determination of the polar angle of~a non-pointing photon with
very good resolution of $0.06/\sqrt{E_{\gamma}/\GeV}$. More over,
the arrival time could be measured with 100\,ps resolution.
This in turn allows for determination of \nino\ decay time $t_D$.
In the Figure~\ref{fig:at-ctauplot} a~correlation between reconstructed
and generated $t_D$ is shown as well as distributions of $t_D/\gamma_{\chi}$
for three values of \ctau.
In the Figure~\ref{fig:at-avedtg} the average 
$\langle t_D/\gamma_{\chi}\rangle$ and
number of lepton non-pointing photon pairs in function of generated
\nino\ \ctau\ is shown after the following selection:
$E_\gamma>30\,\GeV$, $\Delta\theta > 0.2$, $\Delta t_{\gamma} > 1\,\rm ns$
$M_{\rm eff} > 400\,\GeV$, $E^{\rm miss}_{\rm T}> 0.1 M_{\rm eff}$
for integrated luminosity ${\cal L}=13.9/{\rm fb}$.

\begin{figure}
\includegraphics[width=0.5\textwidth,height=0.24\textwidth,angle=0]%
{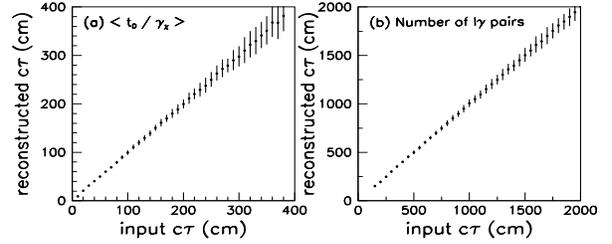}
\caption{Reliability of \ctau\ determination methods.}
\label{fig:at-taures}
\end{figure}

Both values shown in the Figure~\ref{fig:at-avedtg} could be parameterized
and use for \nino\ \ctau\ determination.
The reliability of
both methods and absolute
resolution (error bars) in function of generated \ctau\ is shown
in the Figure~\ref{fig:at-taures}

The relative resolution of \ctau\ ranges
from 3\% to 17\% for
average proper time method $\langle t_D/\gamma_{\chi}\rangle$ and
form 3\% to 6\% for lepton non-pointing $\gamma$ pair counting method.

\section{Non-pointing photons in CMS}
\label{sec:n}

\begin{figure}
\includegraphics[width=.21\textwidth,height=0.21\textwidth,angle=0]%
{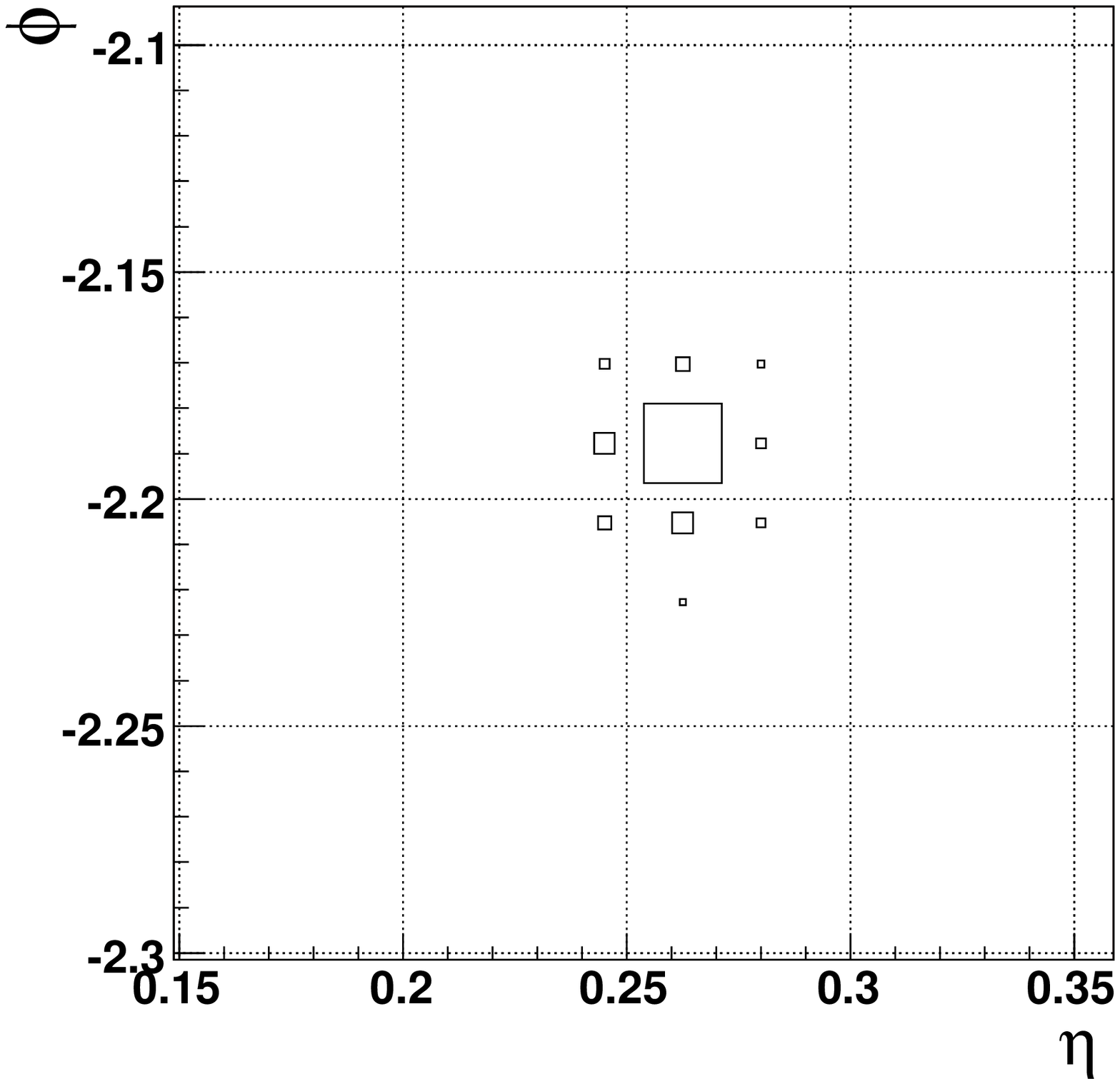}%
\includegraphics[width=.21\textwidth,height=0.21\textwidth,angle=0]%
{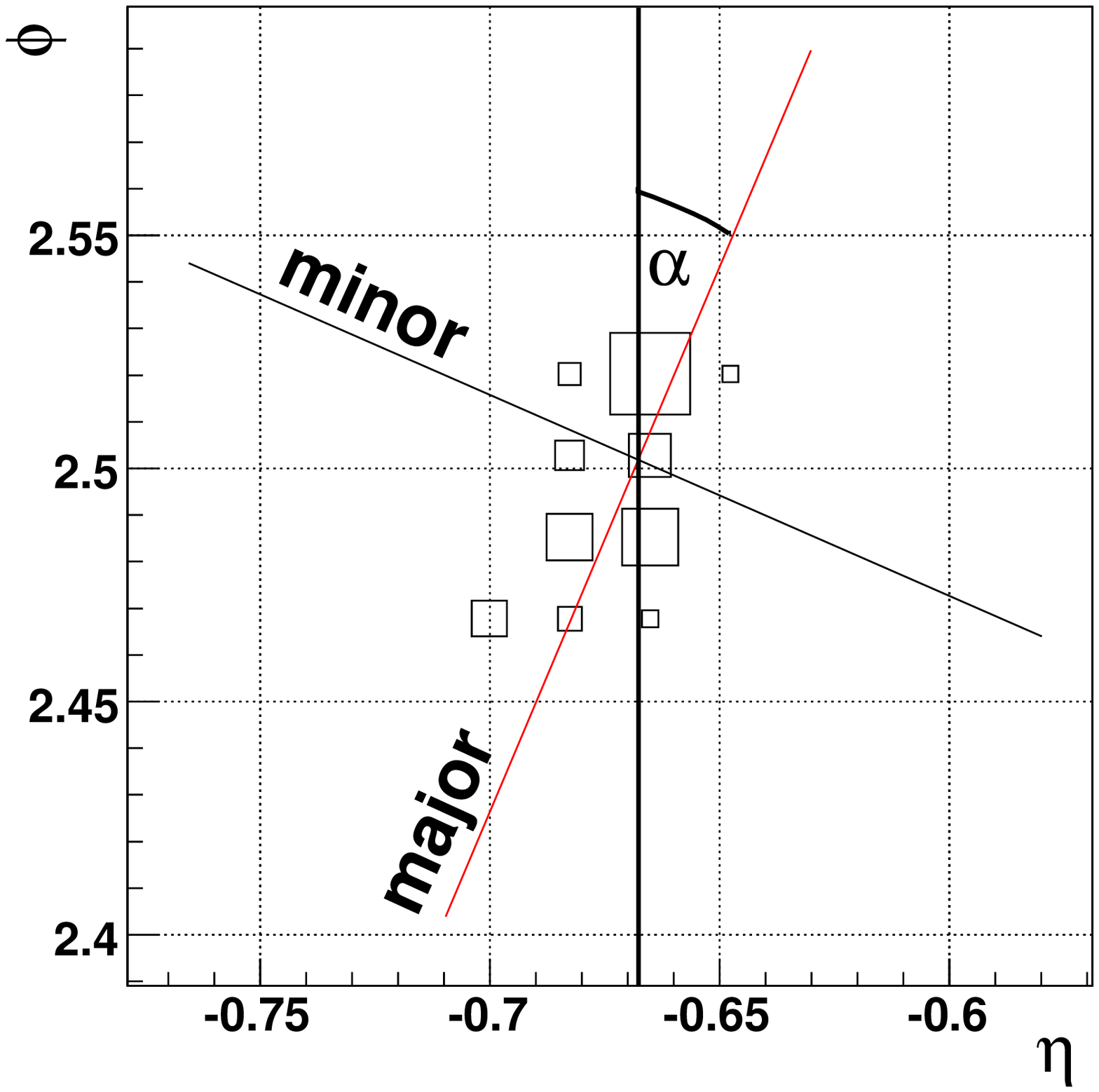}
\caption{Energy distribution in the ECAL crystals for pointing (left)
and non-pointing (right) photon.}
\label{fig:n-alpha}       
\end{figure}

A feasibility study of \nino\ lifetime determination using CMS detector
was done long ago~\cite{ref:zalewski99}. However, only recently and only
part of the original proposal was transformed into full detector simulation 
analysis~\cite{ref:zalewski06}. The method is based on a difference between
shapes of energy distributions among CMS ECAL crystals for pointing 
and non-pointing photons (see Fig.~\ref{fig:n-alpha}).

These energy distributions 
could be characterized by the covariance matrix.
Variances along major and minor axes are given by
$$
{S_{{\rm major}\over{\rm minor}}}
=\frac{S_{\phi\phi}
+S_{\eta\eta}\pm\sqrt{(S_{\phi\phi}-S_{\eta\eta})^2
            +4S_{\phi\eta}^2}}{2}
$$
A measure of the elongation of the deposit is an asymmetry
$$
{\Delta} = \frac{{\Smajor}-{\Sminor}}{{\Smajor}+{\Sminor}} =
 \frac{\sqrt{(S_{\phi\phi}-S_{\eta\eta})^2+4S_{\phi\eta}^2}}
      {S_{\phi\phi}+S_{\eta\eta}}
$$
Another useful variable is the angle $\alpha$ (see Fig.~\ref{fig:n-alpha})
between the major axis
and the $\phi$ axis which could be used to eliminate background due to
converted photons
for which $\alpha\approx0$ and geometrical bias which gives an excess
of deposits with large $\Delta$ for pointing photons for 
$\alpha\approx0,\pm\frac{\pi}{2}$. 

For the full simulation a point from SPS8 line~\cite{ref:SPS78} with
$\Lambda=140\,\TeV$ was chosen and 6 different \nino\ values of \ctau\ were
generated. The \nino\ mass is $192.69\,\GeV$ and the cross section is 0.5\,pb
for this point.

The analysis is based on a search for energetic photons which are accompanied
by at least four hard jets and missing energy, a consequence of cascade
decays of squarks and gluinos. 
A breakdown of the number of events at different 
stages of the selection for signal and background samples is given
in the Table~\ref{tab:n}.

\begin{table}
\caption{%
Number of events at different 
stages of the selection for signal and background for ${\cal L}=10/fb$.
{\bf Preselection}: 
     single $\gamma$ trigger,
     $\pTgst>80\,\GeV$; 
event {\bf selection}: 
     $|\eta^{1j}|<1.7$,
     $\Delta\phi({\rm jj})>20^\circ$,
     $\pTjth>50\,\GeV$,
     $\pTmiss>160\,\GeV$
{\bf non-pointing} $\gamma$ selection:    
     $\Delta>\frac{1}{3}$,
     $\alpha\neq 0,\frac{\pi}{2}$
{\bf pointing} $\gamma$ selection:
     NOT non-pointing,
     $\pTjth>60\,\GeV$,
     $\pTmiss>220\,\GeV$.
}\label{tab:n}
\begin{tabular}{|l|rr|r|r|}
\hline
dataset & {\bf presel.} &  {\bf select.} & {\bf non-p.} &
{\bf point.} \\
\hline
$\ctau=0$
&   2104.28&    402.98&      {2.96}&    {289.39}\\
$\ctau= 25\,\rm cm$
&   2061.69&    379.71&     {29.10}&    {243.76}\\
$\ctau=50\,\rm cm$
&   1948.33&    361.60&     {45.87}&    {215.88}\\
$\ctau=100\,\rm cm$
&   1564.12&    298.12&     {56.03}&    {182.99}\\
$\ctau=200\,\rm cm$
&   1037.60&    166.66&     {34.64}&     {99.39}\\
$\ctau=400\,\rm cm$
&    645.77&    114.89&     {26.16}&     {67.19}\\
\hline
$\Sigma$Zjets&   6137.93&      0.65&      {0.00}&      {0.37}\\
\hline
$\Sigma$Wjets&   8301.93&      2.76&      {0.00}&      {1.46}\\
\hline
$\Sigma$QCD&   787797.19&     54.90&      {0.27}&      {2.32}\\
\hline
VVjets\_incl.
&    836.35&      0.00&      {0.00}&      {0.00}\\
\hline
ttbar\_incl.
&   4662.89&     16.34&      {0.00}&      {6.13}\\
\hline
$\Sigma$total& 807736.31&      74.65&      {0.27}&     {10.27}\\
\hline
\end{tabular}
\end{table}

\begin{figure}
\includegraphics[width=.27\textwidth,height=0.218\textwidth,angle=0]%
{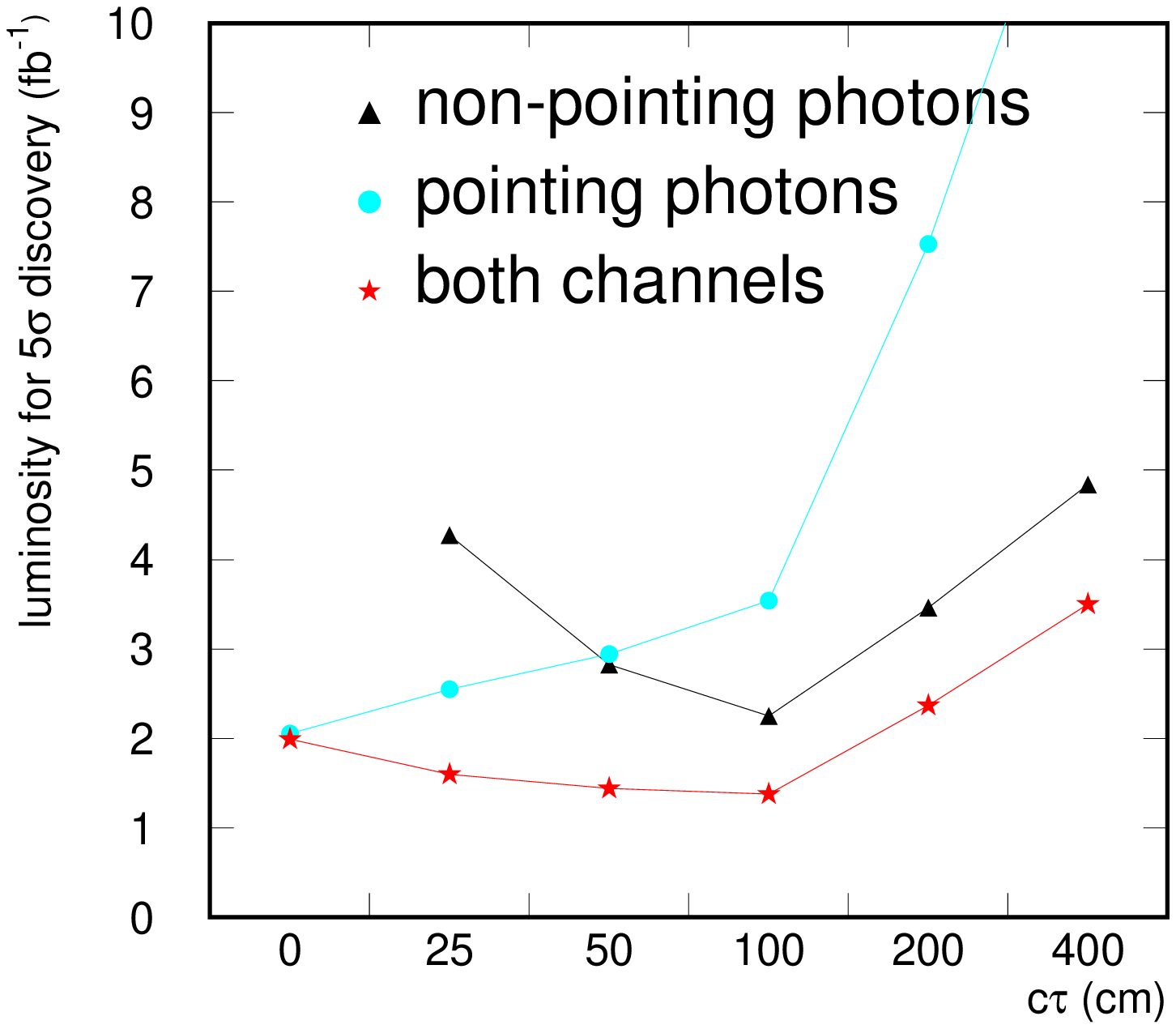}%
\includegraphics[width=.21\textwidth,height=0.21\textwidth,angle=0]%
{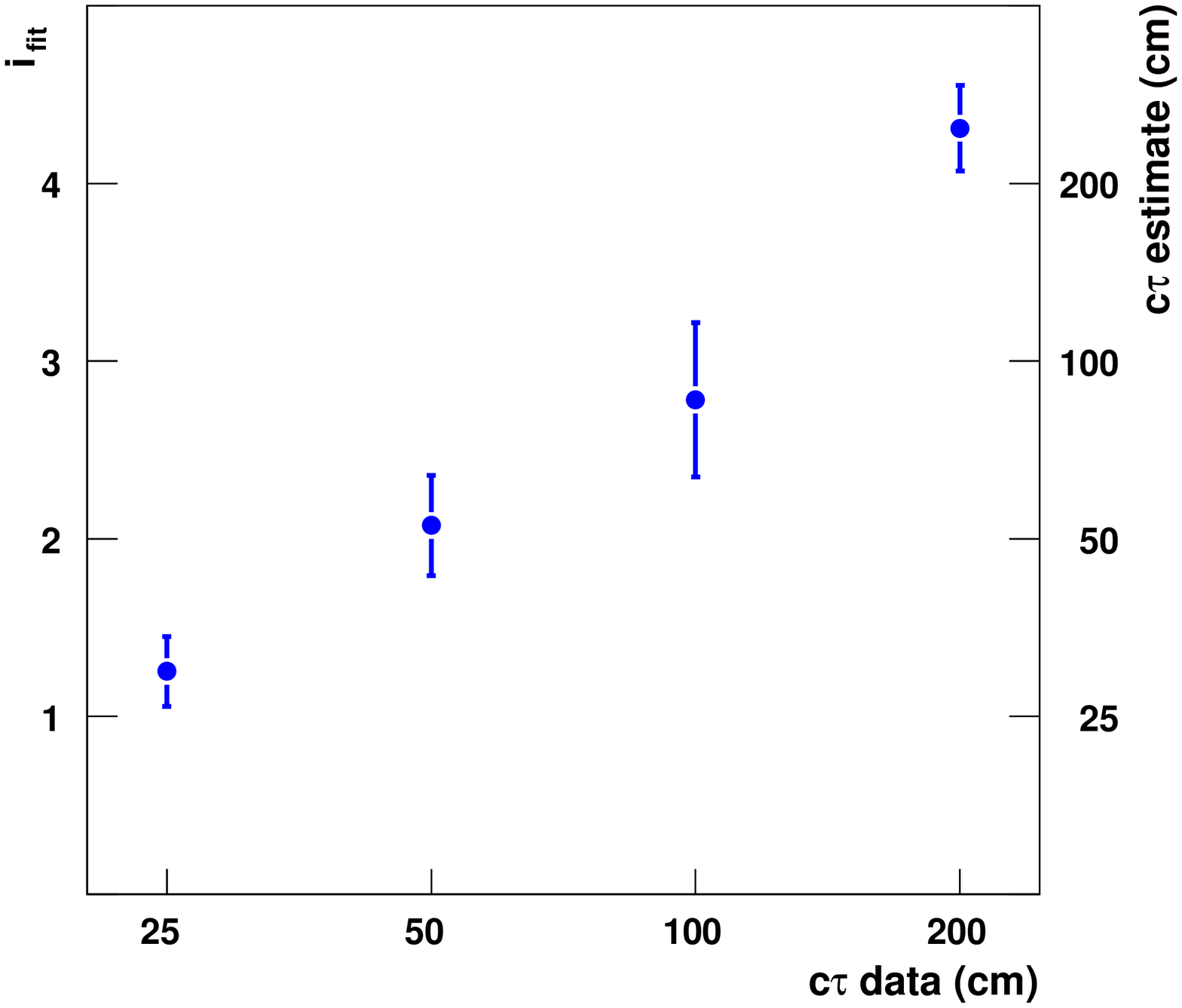}
\mbox{
\begin{minipage}{.2\textwidth}{
\caption{$\cal L$ for $5\sigma$ discovery versus \ctau.}
\label{fig:n-res}}
\end{minipage}\hspace{.05\textwidth}%
\begin{minipage}{.2\textwidth}{
\caption{Estimated \ctau\ versus generated \ctau.}
\label{fig:n-fit}}
\end{minipage}
}
\end{figure}

The fraction of selected photons classified as non-pointing improves
with growing \ctau\ but at the same time more and more neutralinos decays
outside ECAL. Non-pointing selection is almost background free.

An estimate of 
the minimal amount of integrated luminosity needed for $5\sigma$ discovery
versus simulated \ctau\  is
plotted in the Figure~\ref{fig:n-res}.

An integrated luminosity
${\cal L}=2/\rm fb$
is sufficient to claim discovery
for  $\ctau\le100\,\rm cm$
and ${\cal L}=3.5/\rm fb$ for $\ctau=400\,\rm cm$.

To test CMS ECAL capability to estimate \nino\ lifetime
100 likelihood fits of neighborhood \ctau\ and background to the given
\ctau\ + background  was done. The results are shown in the 
Figure~\ref{fig:n-fit}.
The precision of such procedure range from 15\% to 40\%. It will improve
if more \ctau\ points will be used, but one should remember that it
depends on the knowledge of the shapes of distributions of variables
used in the fit. These shapes depend (weakly) on other than \ctau\
parameters of the model.

\section{Conclusions}

Gauge-Mediated Supersymmetry Breaking model is a generic
framework for long-lived, heavy, charged, not strongly interacting particles
and decaying in flight heavy (charged or neutral) particles.

During last decade both Atlas and CMS have invented methods to tackle
with such signatures, despite the fact that detectors had not been
designed for them.


For long lived charged particles both detectors 
use 
synergy of specific ionization measurement
and TOF method (for the full picture see also Ref.~\cite{ref:bressler}). 
The ultimate goal before LHC startup
should be development of model independent methods which could be tested
on cosmics, $Z\ra\mu^+\mu^-$ candle and other sources of energetic muons.

For non-pointing photons signature ATLAS 
and CMS successfully use their electromagnetic calorimeters.
However, the ultimate resolution for lifetime
determination of decaying in flight particles could be obtained only
with tracking systems. Performance of such methods require full
detector simulation with full background turned on.

There is just enough time to do that before LHC startup.

%
%

\end{document}